\documentclass[12pt]{article}
\usepackage{float}
\usepackage{subfloat}
\pdfoutput = 1
\usepackage{graphics}
\usepackage{graphicx} 
\usepackage{amsmath}
\usepackage{amsfonts}   
\usepackage{amssymb}
\usepackage{multirow}
\usepackage{hhline}
\usepackage{color}
\usepackage{comment}

\begin{document}
\title{{\bf{\Large Neutron Star in Quantized-space-time}}}

\author{ {\bf {\normalsize Bhagya. R}\thanks{22phph03@uohyd.ac.in, r.bhagya1999@gmail.com},\
{\bf {\normalsize Diganta Parai}\thanks{digantaparai007@gmail.com}},\,
{\bf {\normalsize E. Harikumar}\thanks{eharikumar@uohyd.ac.in}},\,
{\bf {\normalsize Suman Kumar Panja}\thanks{19phph17@uohyd.ac.in, sumanpanja19@gmail.com}}}\\
{\normalsize School of Physics, University of Hyderabad}\\{\normalsize Central University P.O, Hyderabad-500046, Telangana, India}\\[0.2cm]  
}
\date{}

\maketitle
\begin{abstract}
 We construct and analyze a model of the neutron star in the $\kappa$-deformed space-time. This is done by first deriving the $\kappa$-deformed generalization of the Einstein tensor, starting from the non-commutative generalization of the metric tensor. By generalizing the energy-momentum tensor to the non-commutative space-time and exploiting the $\kappa$-deformed dispersion relation, we then set up Einstein's field equations in the $\kappa$-deformed space-time. As we adopt a realization of the non-commutative coordinates in terms of the commutative coordinates and their derivatives, our model is constructed in terms of commutative variables. Using this, we derive the $\kappa$-deformed generalization of the Tolman–Oppenheimer–Volkoff equation. Now, by treating the interior of the star to be a perfect fluid as in the commutative space-time, we investigate the modification of the neutron star's mass due to non-commutativity of the space-time, valid up to first order in the deformation parameter. We show that the non-commutativity of the space-time enhances the mass limit of the neutron star. We show that the radius and maximum mass of the neutron star depend on the deformation parameter. Further, our study shows that the mass increases as the radius increases for fixed values of the deformation parameter. We show that maximum mass and radius increase as the deformation parameter increases. We find that the mass varies from $0.26M_{\odot}$ to $3.68M_{\odot}$ as radius changes from $8.45km$ to $18.66km$. Using the recent observational limits on the upper bound of the mass of a neutron star, we find the deformation parameter to be $|a|\sim 10^{-44}m$. We also show that the compactness and surface redshift of the neutron star increase with its mass.
\end{abstract}

\section{Introduction}

Supermassive stars are known to become neutron stars, white dwarfs, or black holes towards the last stage of their evolution. The typical core density of these stars is of the order of  $10^{17}\textnormal{kg/m}^{3}$ and the radius is of the order of  $ 10^4m$ \cite{kodama,wheeler1,23}. The Tolman–Oppenheimer–Volkoff (TOV) limit sets an upper bound to the mass of neutron stars. In \cite{24,25}, the TOV limit is estimated to be 2.1 $M_{\odot}$, but one of the recent measurements shows the existence of a neutron star of mass 2.16$M_{\odot}$ \cite{26}. The most massive neutron star observed, PSR $J0740+6620$, is about 2.14 \(M_\odot\)\cite{27}. In \cite{vaidya}, the upper bound on the mass of a neutron star is shown to be 3.575 $M_{\odot}$, using a specific model of space-time metric.

The gravitational force near neutron stars is very high and produces very strong gravitational fields, and thus, it is expected to provide a natural laboratory to test quantum gravity models. One of the approaches in investigating quantum gravity effects advocates that space-time becomes non-commutative \cite{connes} when the gravity is very strong. This approach also incorporates the fundamental length scale seen in several approaches to microscopic gravity. Thus, it is of intrinsic interest to investigate the impact of non-commutativity on neutron stars. 

In this study, we examine neutron star in the $\kappa$-deformed space-time, which is a Lie-algebra type non-commutative space-time (see eq.(\ref{ksp-1})and section 2 for details). Here, we follow the generalization of the approach given in \cite{vaidya} to non-commutative space-time. Non-commutativity has been introduced in our study through the deformed metric and non-commutative generalization of energy-momentum tensor. In this way, we formulate the non-commutative version of Einstein's field and TOV equations. Then, by solving this field equation, we estimate the maximum possible mass a neutron star can have in the $\kappa$-deformed space-time.

Understanding the nature of gravity at the quantum regime is one of the most intriguing topics in physics. Different paradigms are being employed to model and study gravity at microscopic scales \cite{connes, douglas,douglas1, rov, sorkin, Glikman, dop,dop1, am,madore,seiberg-witten}. One characteristic property brought out by all these studies is the defining role of a fundamental length scale in the context of Plank scale gravity. Since the framework of non-commutative geometry has a length scale associated with it, it serves as an environment to construct a model of Plank scale gravity. Thus, modifying general relativity and cosmological models by taking into account the non-commutativity of space-time is of paramount interest.


 In the last couple of decades, extensive studies have been reported on the construction and analysis of different types of non-commutative models \cite {dop,dop1,wess1,wess2,chaichian,chaichian1,kappa1,dimitrijevic,dasz,mel1,mel2,mel3, carlson,amorim}.$\kappa$-deformed space-time is one among these non-commutative space-times where the time and space coordinates obey a Lie algebraic type relation. The associated symmetry algebra has been defined using the Hopf algebra \cite{kappa1,dimitrijevic,dasz,mel1}. The $\kappa$-space-time coordinates satisfy the following commutation relations
\begin{equation}\label{ksp-1}
[\hat{x}^i,\hat{x}^j]=0,~~~[\hat{x}^0, \hat{x}^i]=ia\hat{x}^i,~~~a=\frac{1}{\kappa}.
\end{equation}
In the above, the deformation parameter $a$ has the dimension of length. The non-commutative deformation parameter $a$ encodes the fundamental length scale associated with the quantum gravity effect. The value of $a$ is expected to be close to the Plank length($10^{-35}m$). From the above equation, we recover the commutative space-time in the limit $a\rightarrow 0$.

Recently, various aspects of non-commutative gravity and corresponding physics have been investigated. The effects of the $\kappa$-deformed non-commutativity in cosmology and astrophysics have been analyzed in \cite{zuhair1,zuhair2,visnhu,suman}. In \cite{zuhair1}, $\kappa$-deformed corrections to Hawking radiation are derived using the method of Bogoliubov coefficients. Using $\kappa$-deformed degenerate pressure, compact stars have been studied \cite{zuhair2}. In \cite{visnhu}, the core-envelope model describing superdense stars is constructed using Einstein’s field equation in $\kappa$- deformed space-time. From the $\kappa$-deformed strong energy condition, a bound on the deformation parameter has been obtained. In \cite{suman}, considering space-time to be non-commutative, a detailed investigation of the evolution of the universe within the Newtonian cosmology framework has been discussed. The physics of black holes in non-commutative space-time was analyzed in \cite{kappa-btz,kappa-btz1}. $\kappa$-deformed corrections to the entropy of the BTZ black hole have been calculated using the brick wall method as well as using the quasinormal mode frequency of the $\kappa$-scalar field (in the background of BTZ black hole) \cite{kappa-btz,kappa-btz1}. Non-commutative correction to Bekenstein-Hawking entropy is obtained in \cite{gupta5}. Using a squashed fuzzy sphere, the super Chandrasekhar limit was calculated in \cite{bani}. Investigating the models built upon the framework of general relativity and cosmology in non-commutative space-time will be a good testing ground to see quantum gravity effects and help us understand its nature.

 In \cite{vaidya}, the energy-momentum tensor for perfect fluid distribution in the space-time defined by
\begin{equation}
ds^2=e^{\nu(r)}dt^2-\frac{1-K\frac{r^2}{R^2}}{1-\frac{r^2}{R^2}}dr^2-r^2\left(d\alpha^2+sin^2\alpha d\beta^2\right).
\label{metric}
\end{equation}
 The space part of the space-time is a 3-spheroid embedded in a 4-dimensional Euclidean space. Here $K=1-\frac{l^2}{R^2}$, where $R$ is the equatorial radius of the spheroid, and $l$ is the distance from the center to the north pole along the symmetry axis. In \cite{vaidya}, the space-time metric associated with neutron star is derived to be eq. (\ref{metric}). Einstein's equation, energy-momentum tensor, and TOV equations are derived and solved for this space-time. Using this, it was shown that the upper limit of the mass of the neutron star is $3.575M_{\odot}$ having a radius of $18.37$km. It was shown in \cite{ruff} that the maximum possible mass of a neutron star is $3.2M_{\odot}$. This is obtained by functional maximization procedure, subjected to physical constraints. In the present study, we analyze neutron stars in the $\kappa$-deformed space-time by generalizing the approach of \cite{vaidya}. We construct the $\kappa$-deformed Einstein's equation by deriving the Ricci tensor and Ricci scalar. This is done by using  $\kappa$-deformed metric. By generalizing the energy-momentum tensor to  $\kappa$-deformed space-time, we then derive the $\kappa$-deformed equation of state for the neutron star by solving $\kappa$-deformed Einstein's equation, valid up to the first-order in $a$. Using this, we deduce TOV equations appropriate for the $\kappa$-deformed space-time. We show that the $\kappa$-deformation enhances the mass limit of the neutron star. We show that the upper bound of the neutron star's mass is larger in the non-commutative setting compared to the values obtained for the commutative space-time in \cite{vaidya,ruff}. We also observe that the radius of a neutron star is slightly larger in the $\kappa$-deformed non-commutative space-time compared to the result of the commutative space-time. In particular, the radius of the neutron star is increased by 106m for the minimum allowed value of boundary to core density ratio.

In \cite{garattini}, modification of the TOV equations due to rainbow gravity is analyzed by considering different possible relations between pressure and energy density. Hydrostatic equilibrium conditions for compact stars using the rainbow gravity approach were studied in \cite{hendi}. Here, both radius and mass are functions of the rainbow parameter. It was shown that as maximum mass increases, the radius of the star also increases. But for a fixed value of the rainbow parameter, mass increases as the radius decreases. The effect of the magnetic field on the mass and radius of neutron stars in the rainbow scenario was analyzed in \cite{panah}. It was shown that the maximum mass and radius of the neutron star increase with the magnetic field. In this approach, the maximum mass of the neutron star is argued to be more than 3.2$M_{\odot}$. It is shown that the maximum mass increases when the radius increases. But as in the case of \cite{hendi}, for a fixed value of the rainbow parameter, mass increases as the radius decreases. Incorporating the rainbow functions and non-conserved energy-momentum tensor, modification of the TOV equations for neutron star was derived in \cite{mota}. The effect of various equations of states on the mass was analyzed. In all these cases, mass-radius relations were also obtained. Here, the variation of the mass and radius of the neutron star is derived by fixing the rainbow parameter or the Rastall parameter. For both of these situations, it was shown that the mass decreases as the radius increases. The mass and radius of the dark energy star are investigated in \cite{tudeshki} using the rainbow gravity framework. In this work, the hydrostatic equilibrium condition was implemented through the Chaplygin equation. By varying rainbow function, it was shown that the mass of the star can vary between 2.64 $M_{\odot}$ to 3.7$M_{\odot}$, which is in tune with various observations\cite{abbott1,abbott2,abbott3,thompson,linares,27}. Here, it was shown that as the mass increases from 2.64$M_{\odot}$ to 3.70$M_{\odot}$, the radius of the dark star increases from 12.63 km to 17.69km.

The organization of this paper is as follows. Section 2 represents the essential details of $\kappa$-deformed space-time. In section 3, we show Einstein's equations for the $\kappa$-deformed generalization of the space-time defined by the metric given in eq.(\ref{metric}). Using the $\kappa$-deformed dispersion relation, we then construct the energy-momentum tensor in the deformed space-time. Using these, we set up the $\kappa$-deformed TOV equations. In section 4, we solve these field equations explicitly by considering various physical assumptions. Here, we find the bound on the mass of a neutron star in the $\kappa$-deformed background. In section 5, we present the concluding remarks.

\section{Kappa Deformed Space-Time}

On the $\kappa$-deformed space-time, the field theoretical models were constructed by employing star product formalism \cite{dimitrijevic,dasz}. Alternatively, one could use the realization of non-commutative coordinates. The realization allows one to express non-commutative coordinates in terms of commutative variables \cite{mel1,mel2}. These two approaches are equivalent\cite{mel3}. We adopt the realization approach. We write the $\kappa$-deformed coordinate $\hat{x}_{\mu}$ as \cite{mel1}
\begin{equation}\label{ksp-2}
\begin{split}
 \hat{x}_0=&x_0\psi(A)+iax_j\partial_j\gamma(A)\\
 \hat{x}_i=&x_i\varphi(A).
\end{split}
\end{equation}
here$A=ia\partial_0=ap^{0}$, and $\psi$, $\gamma$, and $\varphi$ are functions of $A$, obeying condition
\begin{equation}\label{ksp-3}
 \psi(0)=1,~\varphi(0)=1.
\end{equation}
Note that eq.(\ref{ksp-2}) is consistent with eq.(\ref{ksp-1}) if $\varphi(A),\varphi^{\prime}(A),\psi(A),\gamma(A)$ satisfy
\begin{equation}\label{ksp-4}
 \frac{\varphi'(A)}{\varphi(A)}\psi(A)=\gamma(A)-1,
\end{equation}
where prime denotes differentiation with respect to $A$. Possible value of $\psi(A)$ are $1$ and $1+2A$  \cite{mel1}. In the present study, we choose $\psi(A)=1$. Thus, equations (\ref{ksp-2}) and  (\ref{ksp-4}) become
\begin{equation}\label{ksp-5}
\begin{split}
 \hat{x}_0=&x_0+iax_j\partial_j\gamma(A)\\
 \hat{x}_i=&x_i\varphi(A),
\end{split}
\end{equation}
and
\begin{equation}\label{ksp-6}
 \frac{\varphi'(A)}{\varphi(A)}=\gamma(A)-1.
\end{equation} 
Here the allowed choices of $\varphi$ are $e^{-A}, e^{-\frac{A}{2}}, 1, \frac{A}{e^A-1}$, etc.  \cite{mel1}. In  \cite{mel1}, it is shown that different choices of $\varphi$ lead to different ordering.
For this realization, the generic form of the free particle dispersion relation is  \cite{mel1}
\begin{equation}
\frac{4}{a^2}\sinh^2 \bigg(\frac{A}{2}\bigg) -p_ip_i \frac{e^{-A}}{\varphi^2(A)}-m^2c^2 +\frac{a^2}{4}\left[\frac{4}{a^2}\sinh^2\bigg(\frac{A}{2}\bigg) -p_ip_i \frac{e^{-A}}{\varphi^2(A)}\right]^2=0.\label{disp}
\end{equation}
The above $p^i$ is the component of the commutative 3-momenta of the particle. Note that in the commutative limit is, i.e., $lim~ a\rightarrow 0$, the above equation reduces to $(p^0)^2-p_i^2-m^2c^2=0$, which is the energy-momentum relation in the commutative space-time. Using realization $\varphi(A)=e^{-A}$, we expand eq.(\ref{disp}) in the powers of deformation parameter and keeping terms up to first order in $a$, we find
\begin{equation}
(p^0)^2-p_i^2(1+ap^0)-m^2c^2=0,
\label{disp_1}
\end{equation}
which we write as
\begin{equation}
p^0=p^0_c+a\tilde{p}^0.
\label{p^0_correction}
\end{equation}
Here $p^0_c$ is the commutative part and $\tilde{p}^0$ is the non-commutative correction. Using this in the eq.(\ref{disp_1}) and comparing the coefficient of the deformation parameter, we get
\begin{eqnarray}
(p^0_c)^2&=&p_i^2+m^2c^2~~~i.e., E^2=p_i^2c^2+m^2c^4\nonumber\\
\tilde{p}^0&=&\frac{1}{2}p_i^2.
\end{eqnarray}
Thus we write
\begin{eqnarray}
p^0=p^0_c+\frac{1}{2}a p_i^2\nonumber\\
i.e., \frac{\hat{E}}{c}=\frac{E}{c}\left(1+\frac{1}{2} a \frac{p_i^2 c}{E}\right).
\label{E_correction}
\end{eqnarray}
where $\hat E$ is the non-commutative energy.
\section{Einstein's field equation in the $\kappa$-deformed space-time}

In this section, we first construct the $\kappa$-deformed metric using the generalized commutation relation between the $\kappa$-deformed phase-space coordinates \cite{zuhair1}. We next obtain the $\kappa$-deformed energy-momentum tensor appropriate for the neutron star. In the $\kappa$-deformed space-time we find Einstein's field equation by promoting  commutative quantities to corresponding $\kappa$-deformed quantities. We start with the generalized  commutation relation for the $\kappa$-deformed phase space coordinates \cite{kappa-geod} as
\begin{equation}\label{N1}
 [\hat{x}_{\mu},\hat{P}_{\nu}]=i\hat{g}_{\mu\nu}, 
\end{equation} 
where $\hat{g}_{\mu\nu}(\hat{x}^{\alpha})$ is the $\kappa$-deformed metric. We choose the $\kappa$-deformed phase-space coordinates as \cite{kappa-geod},
\begin{equation}\label{N2}
 \hat{x}_{\mu}=x_{\alpha}\varphi^{\alpha}_{\mu},~~ \,\hat{P}_{\mu}=g_{\alpha\beta}(\hat{y})p^{\beta}\varphi^{\alpha}_{\mu},
\end{equation}
where $\hat{P}_{\mu}$ is the $\kappa$-deformed generalized momenta corresponding to the non-commutative coordinate $\hat{x}_{\mu}$ and $p_{\mu}$ is the conjugate momenta corresponding to the commutative coordinate $x_{\mu}$. Note here that in the commutative limit, i.e., $a\to 0$, we find $\hat{x}_{\mu}\to x_{\mu}$ and $\hat{P}_{\mu}\to p_{\mu}$. 

Note that in the above equation, we have introduced another set of $\kappa$-deformed space-time coordinates $\hat{y}_{\mu}$.
The coordinates $\hat{y}_{\mu}$ are also assumed to satisfy the $\kappa$-deformed space-time commutation relations  
\begin{equation}
[\hat{y}_0,\hat{y}_i]=ia\hat{y}_i,~~~[\hat{y}_i,\hat{y}_j]=0. \label{N3b}
\end{equation}
Further this $\hat{y}_{\mu}$ is assumed to commute with $\hat{x}_{\mu}$, i.e., $[\hat{y}_{\mu},\hat{x}_{\nu}]=0$. These new coordinates are introduced only for calculational simplification \cite{kappa-geod}. The functional form of $g_{\alpha\beta}(\hat{y})$ in eq.(\ref{N2}) is the same as the metric in the commutative coordinate, but $x_{\mu}$ replaced with non-commutative coordinate $\hat{y}_{\mu}$.

Next we substitute eq.(\ref{N2}) in the $\kappa$-deformed space-time commutation relation, i.e., in eq.(\ref{ksp-1}) and find a particular realisation for $\varphi_{\mu}^{\alpha}$ as
\begin{equation}\label{N3}
 \varphi _0^0=1, \, \varphi _i^0=0, \, \varphi_0^i=0, \, \varphi _j^i=\delta _j^i e^{-ap^0}. 
\end{equation}
Now $\hat{y}_{\mu}$ can be expressed in terms of the commutative coordinates and conjugate momenta as
\begin{equation}\label{N4}
 \hat{y}_{\mu}=x_{\alpha}\phi_{\mu}^{\alpha}.
 \end{equation}
Using eq.(\ref{N3b}) and $[\hat{x}_{\mu},\hat{y}_{\nu}]=0$, one obtains $\phi_{\mu}^{\alpha}$ as (see \cite{zuhair1,kappa-geod} for details)
\begin{equation}\label{N5}
 \phi_{0}^{0}=1,~\phi_{i}^{0}=0,~\phi_{0}^{i}=-ap^i,~\phi_{i}^{j}=\delta_{i}^{j}.
\end{equation}
Thus the explicit form of $\hat{y}_{\mu}$ are
\begin{equation}\label{N6}
 \hat{y}_0=x_0-ax_jp^j,~~
\hat{y}_i=x_i.
 \end{equation}
Using the above in eq.(\ref{N2}) and substituting $\hat{x}_{\mu}$ and $\hat{P}_{\mu}$ in eq.(\ref{N1}), the $\kappa$-deformed metric is obtained as \cite{zuhair1}
\begin{equation}\label{N7}
 [\hat{x}_{\mu},\hat{P}_{\nu}] \equiv i\hat{g}_{\mu\nu}=ig_{\alpha\beta}(\hat{y})\Big(p^{\beta}\frac{\partial \varphi^{\alpha}_{\nu}}{\partial p^{\sigma}}\varphi_{\mu}^{\sigma}+\varphi_{\mu}^{\alpha}\varphi_{\nu}^{\beta}\Big). \end{equation}
Note that $g_{\mu\nu}(\hat{y})$ in the above equation has the same functional form as the commutative metric but is a function of non-commutative coordinates $\hat{y}_{\mu}$. 

Substituting eq.(\ref{N3}) in eq.(\ref{N7}), we find the explicit form of the components of $\hat{g}_{\mu\nu}$ as
\begin{equation}\label{N9}
\begin{aligned}
\hat{g}_{00}&=g_{00}(\hat{y}),\\
\hat{g}_{0i}&=g_{0i}(\hat{y})\big(1-ap^0\big)e^{-ap^0}-ag_{im}(\hat{y})p^me^{-ap^0},\\ 
\hat{g}_{i0}&=g_{i0}(\hat{y})e^{-ap^0},\\
\hat{g}_{ij}&=g_{ij}(\hat{y})e^{-2ap^0}.
\end{aligned}
\end{equation}

Thus the explicit form of the $\kappa$-deformed line element will be \cite{zuhair1}
\begin{equation}\label{N11}
\begin{aligned}
d\hat{s}^2&=g_{00}(\hat{y})dx^0dx^0+\Big(g_{0i}(\hat{y})\big(1-ap^0\big)-ag_{im}(\hat{y})p^m\Big)e^{-2ap^0}dx^0dx^i\\&+g_{i0}(\hat{y})e^{-2ap^0}dx^idx^0+g_{ij}(\hat{y})e^{-4ap^0}dx^idx^j.
\end{aligned}
\end{equation}
From eq.(\ref{N6}) and eq.(\ref{N11}), we observe that the metric components have an explicit dependency on spatial coordinates only. Thus we find $g_{\mu\nu}(\hat{y}^i)=g_{\mu\nu}(x^i)$. Since the cross terms in the metric tensor given in eq.(\ref{metric}) are zero (i.e., $g_{0i}=0,g_{im}=0$), the $\kappa$-deformed metric given in eq.(\ref{N11}) becomes \footnote{Since $\kappa$-deformed space-time is rotational invariant, the $\kappa$-deformed metric is taken to be symmetric in its indices.}
\begin{equation}\label{N12}
 d\hat{s}^2=g_{00}(\hat{y})dx^0dx^0+g_{ij}(\hat{y})e^{-4ap^0}dx^idx^j. 
\end{equation}
 The $\kappa$-deformed space-time metric corresponding to eq.(\ref{metric}) is
\begin{equation}
d\hat{s}^2=e^{\nu(r)}dt^2-e^{-4ap^0}\left[\frac{1-K\frac{r^2}{R^2}}{1-\frac{r^2}{R^2}}dr^2+r^2\left(d\alpha^2+sin^2\alpha d\beta^2\right)\right].
\label{kdeformed_metric}
\end{equation}
Using this we construct the components of the $\kappa$-deformed Ricci tensor($\hat{R}_{\mu\nu}$) and Ricci scalar($\hat{R}$) as follows
\begin{eqnarray}
\hat{R}_{11}=\frac{e^{4ap^0+\nu(r)}}{4r\left(1-K\frac{r^2}{R^2}\right)^2}\Bigg[\left\{4(1+K\frac{r^4}{R^4})-2(K+3)\frac{r^2}{R^2}\right\}\nu^{\prime}(r)+\nonumber\\
r\left(1-\frac{r^2}{R^2}\right)\left(1-K\frac{r^2}{R^2}\right)\left\{\nu^{\prime 2}(r)+2\nu^{\prime\prime}(r)\right\}\Bigg]
\label{R11}
\end{eqnarray}
\begin{equation}
\hat{R}_{22}=\frac{2(1-K)}{R^2}\frac{1+\frac{1}{4}r\nu^{\prime}(r)}{\left(1-\frac{r^2}{R^2}\right)\left(1-K\frac{r^2}{R^2}\right)}-\frac{1}{4}\left\{\nu^{\prime 2}(r)+2\nu^{\prime\prime}(r)\right\}
\label{R22}
\end{equation}
\begin{equation}
\hat{R}_{33}=\frac{r}{2\left(1-K\frac{r^2}{R^2}\right)^2}\left[4r\left(\frac{1-K}{R^2}\right)\left(1-K\frac{r^2}{2R^2}\right)-\left(1-\frac{r^2}{R^2}\right)\left(1-K\frac{r^2}{R^2}\right)\nu^{\prime}(r)\right]
\label{R33}
\end{equation}
\begin{equation}
\hat{R}_{44}=\frac{r sin^{2}\alpha}{2\left(1-K\frac{r^2}{R^2}\right)^2}\left[4r\left(\frac{1-K}{R^2}\right)\left(1-K\frac{r^2}{2R^2}\right)-\left(1-\frac{r^2}{R^2}\right)\left(1-K\frac{r^2}{R^2}\right)\nu^{\prime}(r)\right]
\label{R44}
\end{equation}
\begin{multline}
\hat{R}=\frac{e^{4ap^0}}{2r\left(1-K\frac{r^2}{R^2}\right)^2}\Bigg[2r\left(1-\frac{r^2}{R^2}\right)\left(1-K\frac{r^2}{R^2}\right)\nu^{\prime\prime}(r)+
\Bigg\{4\left(1+K\frac{r^4}{R^4}\right)-
2(K+3)\frac{r^2}{R^2}\Bigg\}\nu^{\prime}(r)\\
+r\left(1-\frac{r^2}{R^2}\right)\left(1-K\frac{r^2}{R^2}\right)\nu^{\prime 2}(r)-
\frac{12(1-K)}{R^2}r\left(1-K\frac{r^2}{3R^2}\right)\Bigg]
\label{R}
\end{multline}
where $\nu^{\prime}(r)$ denotes $\frac{d\nu(r)}{dr}$ and $K$ is the parameter appearing in eq.(\ref{metric}).
Using the eq.(\ref{kdeformed_metric}) and components of $\hat{R}_{\mu\nu}$ and $\hat{R}$ we set up Einstein's tensor in the $\kappa$-deformed space-time. To derive Einstein's field equation, we also need $\kappa$-deformed energy-momentum tensor. For this, we start with the energy-momentum relation valid up to first order in $a$(from eq.(\ref{E_correction})), i.e.,
\begin{equation}
\hat{E}=E\left[1+\frac{1}{2}a p^0\left\{1-\left(\frac{m c^2}{E}\right)^2\right\}\right]\equiv Eg(E).
\label{dispersion_relation}
\end{equation}
The $\kappa$-deformed generalization of the energy-momentum tensor is\footnote{Energy-momentum tensor for a single particle is defined as \cite{wineberg}
\begin{equation}
T^{\alpha\beta}(t,\vec{X})=\frac{p^{\alpha}p^{\beta}}{E}\delta^3\left(\vec{X}-\vec{Y}(t)\right) \nonumber
\end{equation}
 where $p^{\alpha}$ is the four momentum and $\vec{Y(t)}$ is the position of the particle at time $t$.  We extend this definition to non-commutative space-time by replacing all the commutative variables with the corresponding non-commutative variables.}
\begin{equation}
\hat{T}^{\alpha\beta}=\frac{\hat{p}^{\alpha}\hat{p}^{\beta}}{\hat{E}}\delta^3\left(\hat{\vec{X}}-\hat{\vec{Y}}(\hat{t})\right),
\label{energy_momentum_tensor1}
\end{equation}
where $\hat{p}^{\alpha}=m\frac{d\hat{x}^{\alpha}}{d\hat{\tau}}$. For consistency, we demand that $\hat{p}^0=\frac{\hat{E}}{c}$. This condition gives
\begin{equation}
\frac{d \tau}{d \hat{\tau}}=g(E),~~~ \hat{p}^{\alpha}=g(E)\varphi^{\alpha}_{\sigma} p^{\sigma}.
\label{tau_t_rel}
\end{equation}
Using eq.(\ref{energy_momentum_tensor1}),eq.(\ref{tau_t_rel}) and eq.(\ref{N3}) we get
\begin{equation}
\hat{T}^{\alpha\beta}=e^{3ap^0}g(E)\varphi^{\alpha}_{\sigma}\varphi^{\beta}_{\delta}T^{\sigma\delta}.
\label{energy_momentum_tensor2}
\end{equation}
It is known that in the proper frame, $T~is~diag(\rho c^2, P, P, P)$, where $\rho$ and P are the density and pressure of the fluid, respectively. Now using this form of energy-momentum tensor on the RHS of eq.(\ref{energy_momentum_tensor2}), we get
\begin{eqnarray}
\hat{T}^{00}=e^{3ap^0}g(E)\rho c^2\equiv \hat{\rho}c^2\nonumber\\
\hat{T}^{ij}=e^{ap^0}g(E)\delta_{ij}P\equiv \hat{P}\delta_{ij}.
\label{hat_rho_p}
\end{eqnarray}
Thus we find the deformed density and pressure to be $\hat{\rho}=e^{3ap^0}g(E)\rho $ and $\hat{P}=e^{ap^0}g(E)P$, respectively. Using these, we get the general form of the $\kappa$-deformed energy-momentum tensor for a fluid as
\begin{equation}
\hat{T}^{\mu\nu}=\left(\hat{\rho}+\frac{\hat{P}}{c^2}\right)\hat{u}^{\mu}\hat{u}^{\nu}-\hat{P}\hat{g}^{\mu\nu}
\label{energy_momentum_tensor}
\end{equation}
where $\hat{u}^{\mu}=\frac{d\hat{x}^{\mu}}{d\hat{\tau}}$. We derive the explicit form of  $\hat{T}_{\mu\nu}$. Here, we consider the interior of the star to be static perfect fluid, and hence, we take $\hat{u}^{\mu}=(c e^{-\nu(r)/2},0,0,0)$. Thus only diagonal components of eq.(\ref{energy_momentum_tensor}) will survive and these are
\begin{eqnarray}
\hat{T}_{00}&=&c^2 e^{\nu(r)}\hat{\rho}\nonumber\\
\hat{T}_{11}&=&\hat{P}\frac{1-K\frac{r^2}{R^2}}{1-\frac{r^2}{R^2}}e^{-4 a p^{0}}\nonumber\\
\hat{T}_{22}&=&\hat{P}r^2 e^{-4 a p^{0}}\nonumber\\
\hat{T}_{33}&=&\hat{P}r^2 sin^2\alpha e^{-4 a p^{0}}.
\label{component_em_tensor}
\end{eqnarray}
The $\kappa$-deformed Einstein's field equation is given as
\begin{equation}
\frac{8\pi G}{c^4}\hat{T}_{\mu\nu}=\hat{R}_{\mu\nu}-\frac{1}{2}\hat{R}\hat{g}_{\mu\nu}
\label{Einstein_field_equation}
\end{equation}
where $\hat{R}_{\mu\nu}$ and $\hat{R}$ are the $\kappa$-deformed Ricci tensor and Ricci scalar respectively. Using eq.(\ref{R11})-eq.(\ref{R}) and $\hat{T}_{\mu\nu}$ given in eq.(\ref{component_em_tensor}) in eq.(\ref{Einstein_field_equation}) we find
\begin{equation}
\frac{8\pi G}{c^2}\hat{\rho}e^{-4 ap^0}=\frac{3(1-K)}{R^2}\frac{\left(1-K\frac{r^2}{3R^2}\right)}{\left(1-K\frac{r^2}{R^2}\right)^2}
\label{00}
\end{equation}
\begin{equation}
\frac{8\pi G}{c^4}\hat{P}e^{-4ap^0}=\left\{\frac{1-\frac{r^2}{R^2}}{1-K\frac{r^2}{R^2}}\right\}\left[\frac{\nu^{\prime}(r)}{r}+\frac{1}{r^2}\right]-\frac{1}{r^2}
\label{11}
\end{equation}
\begin{eqnarray}
\left(1-\frac{r^2}{R^2}\right)\left(1-K\frac{r^2}{R^2}\right)\left\{\nu^{\prime\prime}(r)+\frac{1}{2}\left[\nu^{\prime}(r)\right]^2-\frac{\nu^{\prime}(r)}{r}\right\}\nonumber\\
-\frac{2(1-K)}{R^2}r\left[\frac{\nu^{\prime}(r)}{2}+\frac{1}{r}\right]+\frac{2(1-K)}{R^2}\left(1-K\frac{r^2}{R^2}\right)=0.
\label{22}
\end{eqnarray}
 Here $\nu^{\prime}(r)=\frac{d\nu}{dr}$ and $\nu^{\prime\prime}(r)=\frac{d^2\nu}{dr^2}$. The eqs. (\ref{00})-(\Ref{22}) is the generalization of the TOV equations to the $\kappa$-deformed space-time with the equation of state given as in eq.(\ref{component_em_tensor}). These equations reduce to the corresponding commutative equations obtained in \cite{vaidya} in the limit $a\rightarrow 0$. At $r=0$ from eq.(\ref{00}) we get
\begin{equation}
\frac{8\pi G}{c^2}\hat{\rho}_{0}e^{-4ap^0}=\frac{3(1-K)}{R^2}
\label{rho_r=0}
\end{equation}
and taking the {\it radius of the star to be} $b$, i.e., on the boundary, where $r=b$, we find
\begin{equation}
\frac{8\pi G}{c^2}\hat{\rho}_{b}e^{-4ap^0}=\frac{3(1-K)}{R^2}\frac{\left(1-K\frac{b^2}{3R^2}\right)}{\left(1-K\frac{b^2}{R^2}\right)^2}~.
\label{rho_r=b}
\end{equation}
Taking derivative with respect to $r$ of eq.(\ref{00})and with the help of eq.(\ref{hat_rho_p}) we obtain
\begin{equation}
\frac{8\pi G}{c^2}g[E]e^{-ap^0}\frac{d\rho}{dr}=\frac{10K(1-K)r}{R^4}\frac{\left(1-\frac{K r^2}{5R^2}\right)}{\left(1-K\frac{r^2}{R^2}\right)^3}~.
\label{derivative_rho}
\end{equation}
From the above expression, it is clear that if we choose $K<0$, then $\rho$ will be a decreasing function of $r$ and always positive. So, from now on, we will consider $K$ to be negative.

Since the ratio of the density of the star at the boundary to that at the center ($\lambda=\frac{\rho_b}{\rho_0}$) is less than one, we find from eq.(\ref{rho_r=0}) and eq.(\ref{rho_r=b})
\begin{eqnarray}
\lambda=\frac{\rho_b}{\rho_0}=\frac{1-K\frac{b^2}{3 R^2}}{\left(1-K\frac{b^2}{R^2}\right)^2}<1.
\label{def_lambda}
\end{eqnarray}
Using the above, we also find
\begin{eqnarray}
\frac{b^2}{R^2}=\frac{1}{6 K \lambda}\left[6\lambda-1-\sqrt{1+24\lambda}\right].
\label{b/R}
\end{eqnarray}
We assume the metric of the exterior region $(r\geqslant b)$ of the star to be  the $\kappa$-deformed Schwarzschild  metric given by
\begin{equation}
ds^2=\left(1-\frac{2 M}{r}\right)dt^2-e^{-4ap^0}\left[\frac{dr^2}{1-\frac{2 M}{r}}+r^2\left(d\alpha^2+sin^2\alpha d\beta^2\right)\right].
\label{Schwarzschild}
\end{equation}
Note that in the notation used, $M$ and $r$ have the same dimension.
Now using the continuity of $g_{11}$(see eq.(\ref{kdeformed_metric}) and eq.(\ref{Schwarzschild}) at $r=b$) we get
\begin{eqnarray}
\left(1-\frac{2 M}{b}\right)^{-1}=\frac{1-K\frac{b^2}{R^2}}{1-\frac{b^2}{R^2}}\nonumber\\
{\rm and}~~~~~~~ M=\frac{(1-K)b^3}{2R^2\left(1-K\frac{b^2}{R^2}\right)}.
\label{mass}
\end{eqnarray}
From eq.(\ref{rho_r=0}) we get
\begin{equation}
R=\sqrt{\frac{3(1-K)c^2}{8\pi G \rho_{0}g(E)}e^{ap^0}}.
\label{radius}
\end{equation}
 Now by specifying $\rho_{b},\lambda,K,ap^0,\frac{mc^2}{E}$, we get the mass and radius of the star from eq.(\ref{mass}) and eq.(\ref{b/R}) with the help of eq.(\ref{radius}). In order to set the physical conditions in
$\kappa$-deformed space-time such as $0<\hat{P}<\frac{1}{3}\hat{\rho}c^2$ and $0<\frac{1}{c^2}\frac{d\hat{P}}{d\hat{\rho}}<1$ throughout the configuration, we need to solve eq.(\ref{11}) and eq.(\ref{22}).

\section{The solution of the field equations}
To find the explicit form of the following metric in eq.(\ref{kdeformed_metric}), we need to solve the eq.(\ref{22}) for $\nu(r)$. For this, we make the change of variables
\begin{eqnarray}
\psi=e^{\nu/2}; ~~ u=\sqrt{\frac{K}{K-1}}\sqrt{1-\frac{r^2}{R^2}}.
\label{variable_change}
\end{eqnarray}
In these new variable eq.(\ref{22}) reduces to 
\begin{equation}
(1-u^2)\frac{d^2\psi}{du^2}+u\frac{d\psi}{du}+(1-K)\psi=0.
\label{change22}
\end{equation}
 We seek a solution in the series form and substitute $\psi=\sum_{n-0}^{\infty}a_{n}u^n$ in the eq.(\ref{change22}). Thus, we get the recursion relation between the coefficients as
\begin{equation}
(n+1)(n+2)a_{n+2}=(n^2-2n+K-1)a_n.
\label{recursion}
\end{equation}
For terminating the series, we choose $K$ that satisfies
\begin{equation}
n^2-2n+K-1=0,~~which~gives~~ n=1\pm\sqrt{2-K}.
\label{condition}
\end{equation}
Since $K$ is negative, the simplest solution is for $K=-2$, i.e., $n=3$. The corresponding solution is
\begin{equation}
e^{\frac{\nu}{2}}=\psi(z)=Bz\left(1-\frac{4}{9}z^2\right)+C\left(1-\frac{2}{3}z^2\right)^\frac{3}{2}
\end{equation}
where we have defined $z=\sqrt{1-\frac{r^2}{R^2}}$. Note that when $r$ ( the distance of a point from the center of a neutron star) is equal to the equatorial radius $R$ of 3-spheroid, $z$ vanishes. $z$ increases from 0 to 1 as $r$ decreases to zero (this $z$ is not redshift parameter). Thus, the final form of the deformed metric is
\begin{eqnarray}
d\hat{s}^2&=&\left\{Bz\left(1-\frac{4}{9}z^2\right)+C\left(1-\frac{2}{3}z^2\right)^\frac{3}{2}\right\}^2dt^2\nonumber\\
&-&e^{-4ap^0}\left[\frac{3-2z^2}{z^2}dr^2+r^2\left(d\alpha^2+sin^2\alpha d\beta^2\right)\right].
\label{final_metric}
\end{eqnarray}
Using solutions for $K$ and $\nu(r)$ in eq.(\ref{00}) and eq.(\ref{11}), we find
\begin{equation}
\frac{8\pi G}{3 c^2}\hat{\rho}=e^{4 ap^0}\frac{5-2 z^2}{R^2\left(3-2 z^2\right)^2}
\label{rho_z}
\end{equation}
\begin{equation}
\frac{8 \pi G}{c^4}\hat{P}=e^{4 ap^0}\frac{3}{R^2}\left\{\frac{C(2z^2-1)\left(1-\frac{2}{3}z^2\right)^\frac{1}{2}-\frac{1}{3}B z(5-4z^2)}{(3-2z^2)\left[C\left(1-\frac{2}{3}z^2\right)^\frac{3}{2}+B z\left(1-\frac{4}{9}z^2\right)\right]}\right\}
\label{P_z}
\end{equation}
Note that eq.(\ref{final_metric}), eq.(\ref{rho_z}) and eq.(\ref{P_z}) are valid in the interior of neutron star. Matching eq.(\ref{Schwarzschild}) and eq.(\ref{final_metric}) at the boundary (i.e at $r=b$) we get
\begin{equation}
\left(1-\frac{2M}{b}\right)=\frac{1-\frac{b^2}{R^2}}{1+2\frac{b^2}{R^2}}
\label{g11}
\end{equation}
\begin{equation}
B\left(1-\frac{b^2}{R^2}\right)\left(5+4\frac{b^2}{R^2}\right)+C\sqrt{3}\left(1+2\frac{b^2}{R^2}\right)^{\frac{3}{2}}=9\left(1-\frac{2M}{b}\right)^{\frac{1}{2}}.
\label{g00}
\end{equation}
Now we demand that the fluid pressure must vanish at the boundary \cite{vaidya} and using the definition of $z$ in eq.(\ref{P_z}), we obtain
\begin{equation}
B\left(1-\frac{b^2}{R^2}\right)^\frac{1}{2}\left(1+4\frac{b^2}{R^2}\right)=C\sqrt{3}\left(1-2\frac{b^2}{R^2}\right)\left(1+2\frac{b^2}{R^2}\right)^\frac{1}{2}.
\label{pressure_condition}
\end{equation}
By solving eq.(\ref{g11}), eq.(\ref{g00}) and eq.(\ref{pressure_condition}) we find
\begin{eqnarray}
B&=&\frac{3}{2}\frac{1-2\frac{b^2}{R^2}}{\left(1+2\frac{b^2}{R^2}\right)^\frac{1}{2}}\nonumber\\
C&=&\frac{\sqrt{3}}{2}\sqrt{1-\frac{b^2}{R^2}}\left\{\frac{1+4\frac{b^2}{R^2}}{1+2\frac{b^2}{R^2}}\right\}
\label{AB}
\end{eqnarray}
By dividing eq.(\ref{P_z}) by eq.(\ref{rho_z}) we get
\begin{equation}
\frac{\hat{P}}{\frac{1}{3}\hat{\rho}c^2}=\frac{3\left(3-2 z^2\right)}{5-2 z^2}\left[\frac{C(2z^2-1)\left(1-\frac{2}{3}z^2\right)^\frac{1}{2}-\frac{1}{3}B z(5-4z^2)}{C\left(1-\frac{2}{3}z^2\right)^\frac{3}{2}+B z\left(1-\frac{4}{9}z^2\right)}\right].
\label{P/rho}
\end{equation}
By generalizing the requirement of strong energy condition to the $\kappa$-deformed situation, we set
\begin{equation}
\hat{P}<\frac{1}{3}\hat{\rho}c^2.
\label{strong_energy}
\end{equation}
Using the above condition( i.e.$0<\frac{\hat{P}}{\frac{1}{3}\hat{\rho}c^2}<1$) and with the help of eq.(\ref{P/rho}) and eq.(\ref{AB}) we get a bound on the value of $\frac{b^2}{R^2}$(see the table-\ref{tab1}). We emphasize that the $z$ used here is not the redshift parameter. $z$ ranges from 1 to 0 as one moves out from the center of the neutron star to the edge of the 3-spheroid. By choosing different values of $z$ and imposing the strong energy condition given (in eq.(\ref{strong_energy})) (\ref{P/rho}), we find the allowed values of $\frac{b^2}{R^2}$, where $b$ is the radius of the neutron star.

\begin{table}[h!]
\caption{Condition on $\frac{b^2}{R^2}$ for different values of $z=\sqrt{1-\frac{r^2}{R^2}}$ satisfying strong energy condition.\label{tab1}}
    \centering
     \begin{tabular}{|c|c|}
    \hline
    \textbf{ $z=\sqrt{1-\frac{r^2}{R^2}}$}&\textbf{Condition on the value of $\frac{b^2}{R^2}$ due to constraint $0<\frac{\hat{P}}{\frac{1}{3}\hat{\rho}c^2}<1$}  \\
    \hline
    1 & 0$<\frac{b^2}{R^2}<$0.3167  \\
    \hline
    0.95 & 0.0975$<\frac{b^2}{R^2}<$0.3708  \\
    \hline
    0.9 & 0.19$<\frac{b^2}{R^2}<$0.4278  \\
    \hline
    0.85 & 0.2775$<\frac{b^2}{R^2}<$0.485  \\
    \hline
    $\sqrt{1-0.3167}=0.8266$ & 0.3167$<\frac{b^2}{R^2}<$0.5113 \\
    \hline
    0.8 &0.36$<\frac{b^2}{R^2}<$0.5407  \\
    \hline
    \end{tabular}
    
\end{table}
From table-\ref{tab1}, it is evident that to satisfy the strong energy condition at every point inside the star, the upper bound on $\ \frac {b^2}{R^2}$ is $0.3167$. We see from eq.(\ref{AB}) that within this bound on $\frac{b^2}{R^2}$, $B$~and$C$  cannot be negative. So, for physical configurations, $\frac{B}{C}$ will always be positive. Another condition is that the speed of sound should be less than the speed of light within the configuration (as required by causality). By definition, $\frac{dP}{d\rho}$ is the square of sound velocity. Thus, in the $\kappa$-deformed space-time causality condition will be
\begin{equation}
\frac{d\hat{P}}{d\hat{\rho}}<c^2.
\end{equation}
By taking the derivatives  of eq.(\ref{rho_z}) and eq.(\ref{P_z}) with respect to $z$, and using them we get
\begin{equation}
\frac{1}{c^2}\frac{d\hat{P}}{d\hat{\rho}}=\frac{1}{\frac{d}{dz}\left\{\frac{5-2 z^2}{\left(3-2 z^2\right)^2}\right\}}\frac{d}{dz}\left\{\frac{(2z^2-1)\left(1-\frac{2}{3}z^2\right)^\frac{1}{2}-\frac{1}{3}\frac{B}{C} z(5-4z^2)}{(3-2z^2)\left[\left(1-\frac{2}{3}z^2\right)^\frac{3}{2}+\frac{B}{C} z\left(1-\frac{4}{9}z^2\right)\right]}\right\}
\end{equation}

\begin{table}[h!]
\caption{Condition on the value of $\frac{B}{C}$ for different values of $z=\sqrt{1-\frac{r^2}{R^2}}$ to satisfy the causality condition.\label{tab2}}
\centering   
\begin{tabular}{|c|c|}
\hline
\textbf{$z=\sqrt{1-\frac{r^2}{R^2}}$}& \textbf{Condition on the value of $\frac{B}{C}\equiv  x$ due to constraint $0<\frac{1}{c^2}\frac{d\hat{P}}{d\hat{\rho}}<1$} \\
    \hline
    1 & $x>0.1762$  \\
      \hline
    0.98 & $x>0.1765$ \\
      \hline
    0.96 & $x>0.1741$  \\
      \hline
    0.94 & $x>0.1691$  \\
      \hline
    0.92 & $x>0.1619$ \\
      \hline
    0.90 & $x>0.1527$ \\
     \hline
    0.88 & $x>0.1417$ \\
      \hline
   \end{tabular}
 \end{table}
From table-\ref{tab2}, we observe that to satisfy the causality condition, the lower bound on $\frac{B}{C}$ should be $0.1762$. Now, we find the expression for the mass of the neutron star valid up to the first order in the deformation parameter. Using eq.(\ref{dispersion_relation}) in eq.(\ref{radius}) we get the expression of $R$ 
\begin{equation}
R=R_0\left[1+\frac{1}{2}\alpha ap^0\right]~~where~~ R_{0}=\sqrt{\frac{9c^2}{8\pi G \rho_{0}}},~~and~~\alpha\equiv \left[1-\frac{1}{2}\left\{1-\left(\frac{m c^2}{E}\right)^2\right\}\right].
\label{radius_final}
\end{equation}
Note that we have taken correction term valid up to the first order in the deformation parameter. From eq.(\ref{b/R}) we get the {\it radius of the star} to be
\begin{equation}
b=b_{0}\left[1+\frac{1}{2}\alpha ap^0\right],~~where~~b_0=\sqrt{\frac{R_0^2}{12\lambda}\left[1+\sqrt{1+24\lambda}-6\lambda\right]},
\label{b_final}
\end{equation}
and the expression of $\lambda$ is given in eq.(\ref{def_lambda}).
Using eq.(\ref{radius_final}), eq.(\ref{b_final}) and eq.(\ref{mass}) we  get
\begin{equation}
M=\frac{3b_0^3}{2R_0^2\left(1+2\frac{b_0^2}{R_0^2}\right)}\left(1+\frac{1}{2}\alpha ap^0\right)=M_{0}\left(1+\frac{1}{2}\alpha ap^0\right),
\label{m_final}
\end{equation}
where $M_0$ is the mass of the neutron star in the commutative case.
As we expect the deformation parameter to be of the order of Planck length, we set $a=10^{-35}$ meter and consider the value of $ap^0=0.01$.
Considering $p^0(=\frac{E}{c})$ to be $10^6M_{\odot}$ ( the mass of the black hole) we find $\frac{mc^2}{E}\sim 10^{-63}$. Using this in eq.(\ref{radius_final}) we obtain $\alpha\approx 0.5$,
where $m$ is the mass of the neutron. We take $\rho_b=2\times 10^{17}kg~m^{-3}$\cite{vaidya} to calculate $R$ for different choices of $\lambda$. Below, we present the ratios of the masses of neutron stars to the sun for various values of the parameter. In our calculations, we use the mass of the sun($M_{\odot}$) to be 1475 meters (in natural units).  Compactness of an object is defined as the ratio of its  mass to radius \cite{visnhu,kalam,maurya} \footnote{Note that this definition differs from the one used in \cite{hendi} by an overall multiplicative factor of 2.}
\begin{equation}
u=\frac{M}{b}.
\end{equation}
Compactness is an indicator of the gravitational strength of objects. From eq.(\ref{m_final},\ref{b_final}) we find that the $u$ defined in the above equation independent of the deformation parameter (up to first order). Explicitly, we get
\begin{equation}
u=\frac{M}{b}=\frac{3b^2}{2 R^2\left(1+\frac{2b^2}{R^2}\right)}.
\end{equation}
Using the compactness, one calculates the surface redshift \cite{kalam,maurya} as
\begin{equation}
Z_{redshift}=\frac{1}{\sqrt{1-2 u}}-1=\sqrt{\frac{1+2\frac{b^2}{u^2}}{1-\frac{b^2}{R^2}}}-1.
\end{equation}
which is also a good indicator of the strength of the gravitational field produced by objects. Note here that the surface redshift is also independent of the non-commutative parameter (up to first  order).
For different values of $\lambda$, we have calculated compactness and surface redshift (see table\ref{tab3}).
\begin{table}[h!]
\caption{ The mass and radius of neutron star. \label{tab3} }
\centering
\begin{tabular}{|c|c|c|c|c|c|}
\hline
\textbf{$\lambda$}&\textbf{$\left(\frac{b}{R}\right)^2$}&\textbf{$b(radius~~of~~star)(km)$}&\textbf{$\frac{M}{M_{\odot}}$}& compactness & $Z_{redshift}$\\
\hline
    0.9 &0.0327&8.45&0.2644 &0.046 &0.0495\\ 
    \hline
   0.8&0.0723&11.84&0.7610 &0.0947 &0.1107\\
   \hline
   0.7&0.1213&14.34&1.4241 &0.1464 &0.1892\\
   \hline
   0.6&0.1839&16.35&2.2361 &0.2017 &0.2946\\
   \hline
   0.5&0.2676&18.01&3.1917 &0.2615 &0.4478\\
   \hline
   0.4539&0.3167&18.66&3.6805 &0.2908 & 0.5461\\
   \hline
   0.4&0.3866&19.35&4.2921 &0.3270 &0.7\\
   \hline
  
   \end{tabular}
\end{table}

Note that the compactness and surface redshift are increasing with the mass of the neutron star.Similar feature was reported for neutron stars in rainbow gravity also\cite{hendi}. From the mass-radius plot given below (Figure 1), we see that the radius of the star increases with its mass.

\begin{figure}[h!]
\centering
\includegraphics[scale=1.2]{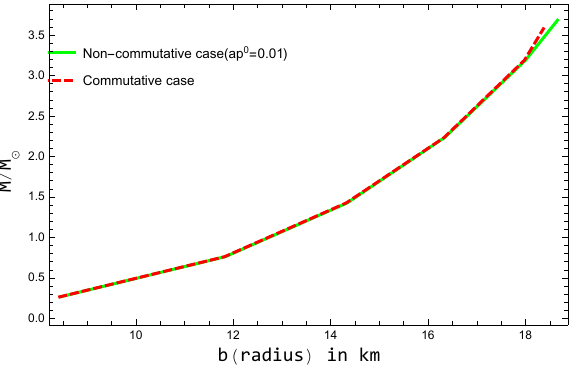}
\caption{$M/M_{\odot}$ vs $b$(radius)}
\end{figure}

Note that $\lambda\leq0.4$ is ruled out by the condition on $\left(\frac{b}{R}\right)^2$ (see table-\ref{tab1}).
From eq.(\ref{m_final}), we observe that, due to the $\kappa$-deformed correction, the mass of the neutron star is larger than the result reported in \cite{vaidya} for the commutative case. For fixed $ap^0=0.01$, we find that as the radius of the neutron star increases, mass also increases. We tabulate the maximum of $\frac{M}{M_{\odot}}$ ratio as a function of $ap^{0}$ in the table-\ref{tab4} below. From table-\ref{tab4}, we see that, as we increase the value of $ap^0$, the maximum mass of the neutron star (in solar mass unit) and radius increase. The plot below(figure-2) makes it clear that the maximum value of $\frac{M}{M_{\odot}}$ increases linearly with the deformation function $ap^0$.

\begin{table}[h!]
\caption{Maximum mass of neutron star for different values of $ap^{0}$ .\label{tab4}}
\centering
\begin{tabular}{|c|c|c|}
\hline
    \textbf{$ap^{0}$} & \textbf{Maximum value of $\frac{M}{M_{\odot}}$ for neutron star}& b (radius in km) \\
\hline
    0.01 & 3.6805 & 18.665  \\
    \hline
    0.02 & 3.6897 & 18.711 \\
    \hline
    0.03 & 3.6988 & 18.758  \\
    \hline
    0.04 & 3.7080 & 18.804  \\
    \hline
    0.05 & 3.7172 & 18.851 \\
    \hline
    0.06 & 3.7264 & 18.897 \\
    \hline
    0.07 & 3.7356 & 18.944\\
    \hline
    0.08 & 3.7447 & 18.991\\
    \hline
    0.09 & 3.7539 & 19.037\\
    \hline
     0.1 & 3.7631 & 19.084\\
   \hline
    \end{tabular}
 \end{table}

\begin{figure}[h!]
\centering
\includegraphics[scale=1.2]{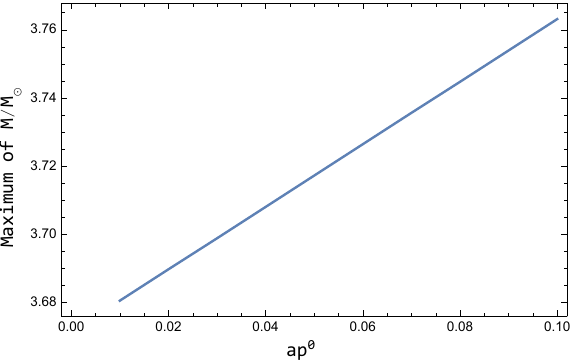} 
\caption{$M_{max}/M_{\odot}$ vs $ap^0$}
\end{figure}

Recent observations of PSR $J0740+6620$ set the bound on the mass of neutron star to be $2.14M_{\odot}$ \cite{27}. Analysis of mass-tidal deformation data from  PSR $J0030+0451$ put this value to be $2.25^{+0.08}_{-0.07}M_{\odot}$. Using these values on the left hand side of eq.(\ref{m_final}), we find $|\alpha ap^{0}|$ to be $0.8$ and $0.7$ respectively (or $|a|\sim 10^{-44}m$).

\section{Conclusions}
Compact objects such as white dwarfs, neutron stars, and black holes have high mass density and produce extremely strong gravitational fields. Since very strong gravitational fields are expected to modify the space-time structure, neutron stars provide testing grounds to study signals for quantum gravity. Here, we constructed a model of the neutron star in the $\kappa$-deformed space-time, a non-commutative space-time that is relevant for quantum gravity models. We showed that the non-commutativity of space-time increases the upper bound on the mass of a neutron star. For $ap^0=0.01$, we have found that the upper limit of neutron star mass is $3.6805M_{\odot}$. Recent analysis of observational data put this bound to be 2.14 $M_{\odot}$ \cite{27} and mass-tidal deformation data sets it to be $2.25^{+0.08}_{-0.07}$ $M_{\odot}$ \cite{fan}. We find that these values set the values of the deformation parameter $a$ to be  $-0.27\times 10^{-44}m$ and $-0.25\times 10^{-44}m$, respectively. We found that for $ap^0=0.01$, the change in radius of the maximum mass neutron star is around 3 percent larger than the commutative result. For the astronomical length scale, this change is small.

 After generalizing the metric eq.(\ref{metric}) to the non-commutative space-time and deriving a deformed energy-momentum tensor, we constructed the modified Einstein's equation describing the neutron star. The solution of deformed TOV equations is derived and valid up to the first order in the deformation parameter. We have exploited the $\kappa$-deformed strong energy condition and causality condition in obtaining the solution of the modified TOV equation.

Various studies related to the upper bound on the mass of a neutron star have been reported in literature \cite{24,25,26,27, vaidya, fan, ruff}. In \cite{24,25}, the upper bound on the mass of a neutron star is found to be 2.1 $M_{\odot}$. It is estimated to be 2.16 $M_{\odot}$ in a recent paper \cite{26}. Using observational data, the maximum mass of a neutron star is found to be 2.14 $M_{\odot}$ for PSR $J0740+6620$ in \cite{27}. Maximum gravitational mass of a neutron star is obtained to be $2.25^{+0.08}_{-0.07}$ $M_{\odot}$ using the mass-tidal de-formability data of GW170817 and the mass-radius data of PSR $J0030+0451$ and PSR $J0740+6620$ \cite{fan}. From a purely geometrical point of view, the upper bound on the mass of a neutron star is found to be 3.575 $M_{\odot}$ \cite{vaidya}. In \cite{25}, it has been concluded that the true maximum mass of the neutron star is between $2M_{\odot}$ and $3M_{\odot}$. In \cite{ruff}, it is argued that the mass of a neutron star with zero angular momentum can not be more than 3.2 $M_{\odot}$ for any equation of state at nuclear density. 

In our study, we find that the upper bound for the mass of the neutron star varies with the non-commutative scale. Here, we have considered the matter density on the star boundary (i.e.$r=b$) to be $\rho=2\times 10^{17}kg~m^{-3}$ \cite{vaidya}. Various possible values for the ratio $\lambda=\frac{\rho_b}{\rho_0}$ have been considered (where $\rho_0$ is the density at the center of the star). By specifying $ap^0$ from eq.(\ref{radius_final}) we find $R$. Using $R$, from eq.(\ref{b_final}) we find the radius of the star($b$). Finally, from eq.(\ref{m_final}) mass of the star is found. For a fixed value of $ap^0=0.01$ and various values of $\lambda$, we present the radius and mass of the neutron star(as a multiple of solar mass $M_{\odot}$) in table-\ref{tab3}. $\kappa$-deformed strong energy condition indicates that the ratio of the boundary density to the core density of a neutron star ($\lambda$) can not be less than $0.4539$ for a neutron star model. The first six values of $\lambda$ in table-\ref{tab3} correspond to the physically viable star model. For our specific choice of  $ap^0=0.01$, we found that the maximum mass of a neutron star is $3.6805M_{\odot}$ having a radius $18.66$ km, which is higher than the result reported in\cite{vaidya, ruff}. We have shown that the compactness and surface redshift increase with the mass of the  neutron star in $\kappa$-deformed space-time. It is seen from table-\ref{tab4} that if we increase the value of the deformation parameter, the mass limit will enhance. Thus, we see that the effect of space-time non-commutativity increases the limiting mass of a neutron star. In \cite{tudeshki}, it was shown that as the mass of the neutron star increases, the radius also increases. This feature is the same when the rainbow function is fixed or varied. In \cite{hendi,panah, mota}, as the maximum mass increases, the radius also increases when the rainbow parameter is varied. But in all these studies for a fixed value of the rainbow parameter, it is shown that the radius decreases as the mass increases. Our study exhibits the feature that the mass of the neutron star increases as the radius increases. This behavior is the same for fixed deformation parameters and when the deformation parameter increases. This behavior of mass-radius relation in the $\kappa$-deformed space-time is similar to the one reported in \cite{tudeshki}. But the feature of increase in the radius as a mass of neutron star increases for fixed deformation parameter is in contrast to the result obtained for the fixed rainbow parameter in \cite{hendi,panah, mota}. It is reported from the observation that the average radius of the neutron star is around 10km\cite{golden}. There is a theoretical prediction that the mass of a neutron star can be above 3$M_{\odot}$. In the $\kappa$-deformed space-time, the neutron star mass is 3.5$M_{\odot}$ and radius (b) to be 10km with $ap^0=0.01$ we find the density of the center of the neutron star to be $5.28\times 10^{18}kg~m^{-3}$.

Our study shows that the maximum mass of the neutron star is enhanced by the non-commutativity of the space-time. This provides us with a possible way to account for neutron stars with larger masses. We also see that the Planck scale modification of the structure of the space-time fabric can have observable signals at large length scales. Different frameworks of studying quantum gravity effects in analyzing neutron stars have been reported\cite{garattini,hendi,panah, mota,tudeshki}. It is fascinating to see whether the observed values of the neutron star parameters can be used to select the preferred quantum gravity model from these.

\section{Acknowledgement}
BR thanks DST-INSPIRE for support through the INSPIRE fellowship (IF220179). DP thanks  IOE-UOH for support through the PDRF scheme. S.K.P thanks UGC, India, for the support through the JRF scheme (id.191620059604).


\end{document}